\documentclass[letterpaper, 10 pt, conference]{ieeeconf}  

\IEEEoverridecommandlockouts                              
\usepackage{tikz}

\usepackage{amsthm}    
\usepackage{amsmath} 
\usepackage{amssymb}  
\usepackage{mathtools}
\usepackage{algpseudocode}
\usepackage{algorithm}
\usepackage{multirow}

\DeclareMathOperator{\diag}{diag}
\DeclareMathOperator{\blkdiag}{blkdiag}
\newtheorem{remark}{Remark}
\newtheorem{theorem}{Theorem}
\newtheorem{corollary}{Corollary}
\newtheorem{definition}{Definition}
\newtheorem{lemma}{Lemma}
\newtheorem{problem}{Problem}

\newcommand{\bbC}{\mathbb{C}}

\newcommand{\bbN}{\mathbb{N}}

\newcommand{\bbR}{\mathbb{R}}
\newcommand{\bbS}{\mathbb{S}}

\newcommand{\calE}{\mathcal{E}}

\newcommand{\diff}{\,\mathrm{d}}

\DeclareMathOperator{\slope}{slope}
\DeclareMathOperator{\trace}{trace}
\newcommand{\ovl}{\overline}
\newcommand{\udl}{\underline}

\title{\LARGE \bf
Linear systems with neural network nonlinearities: Improved stability analysis via acausal Zames-Falb multipliers}

\author{Patricia Pauli, Dennis Gramlich, Julian Berberich and Frank Allgöwer
\thanks{This work was funded by Deutsche Forschungsgemeinschaft (DFG,
German Research Foundation) under Germany’s Excellence Strategy - EXC
2075 - 390740016. We acknowledge the support by the Stuttgart Center for Simulation Science (SimTech). The authors thank the International Max Planck Research
School for Intelligent Systems (IMPRS-IS) for supporting Patricia Pauli, Dennis Gramlich and Julian Berberich.}
\thanks{Patricia Pauli, Dennis Gramlich, Julian Berberich and Frank Allg\"ower are with the Institute for Systems Theory and Automatic Control, University of Stuttgart, 70569 Stuttgart, Germany. Dennis Gramlich is with the Chair of Intelligent Control Systems, RWTH Aachen University, 52074 Aachen, Germany. E-Mail:
{\tt\small patricia.pauli@ist.uni-stuttgart.de}}%
}

\begin{document}

\maketitle
\thispagestyle{empty}
\pagestyle{empty}

\begin{abstract}

In this paper, we analyze the stability of feedback interconnections of a linear time-invariant system with a neural network nonlinearity in discrete time. Our analysis is based on abstracting neural networks using integral quadratic constraints (IQCs), exploiting the sector-bounded and slope-restricted structure of the underlying activation functions. In contrast to existing approaches, we leverage the full potential of dynamic IQCs to describe the nonlinear activation functions in a less conservative fashion. To be precise, we consider multipliers based on the full-block Yakubovich / circle criterion in combination with acausal Zames-Falb multipliers, leading to linear matrix inequality based stability certificates. Our approach provides a flexible and versatile framework for stability analysis of feedback interconnections with neural network nonlinearities, allowing to trade off computational efficiency and conservatism. Finally, we provide numerical examples that demonstrate the applicability of the proposed framework and the achievable improvements over previous approaches. 

\end{abstract}

\section{Introduction}
Deep neural networks (NNs) are a powerful, efficiently trainable, and broadly applicable tool to describe nonlinear input-output behavior. However, safety certificates are urgently required in order to enable the use of deep NNs in safety-critical applications such as autonomous driving. Lately, along with new advances in deep learning, there has been an increasing interest in the analysis of deep NNs using quadratic constraints \cite{fazlyab2020safety,hashemi2021certifying}, originally suggested in \cite{levin1993control,suykens1995artificial}. Aiming at robustness and stability guarantees for NNs, this branch of research has for example been concerned with estimating the Lipschitz constant of NNs \cite{fazlyab2019efficient,pauli2021training} and giving closed-loop stability certificates for NN controllers \cite{yin2020stability,pauli2020offset}.

In this paper, we study stability of the feedback interconnection of a linear time-invariant (LTI) system and an NN nonlinearity in discrete time.
Within this setup, the NN can represent a controller that, e.g., results from a reinforcement learning algorithm or is trained to approximate a computationally more expensive model predictive controller (MPC). Our approach may then be used to verify stability of the resulting feedback interconnection. Alternatively, the NN may result from a nonlinear system identification step, in which case stability properties of the identified model may be sought. The setup hence contains recurrent NNs as a special case whose analysis has been addressed using incremental quadratic constraints in \cite{revay2020convex}. 

For our analysis, we leverage integral quadratic constraints (IQCs) \cite{megretski1997system, veenman2016robust} to abstract NNs by exploiting that the most common activation functions are memoryless, sector-bounded, slope-restricted nonlinearities. The feedback interconnection of an LTI system with a fully-connected feed-forward NN is a Lur'e system, where the nonlinearities within the NN structure are isolated via a loop transformation. Stability of the system can then be analyzed using Lyapunov function arguments, yielding a linear matrix inequality (LMI) certificate for stability \cite{gonzaga2012stability,ahmad2012lmi}.

In this paper, we extend the local stability analysis established in \cite{yin2020stability} to the use of more general IQCs.
Previous works that abstract NNs using quadratic constraints \cite{fazlyab2019efficient,pauli2021training} employ static IQCs to capture the slope-restricted nature of the activation functions. To the best of our knowledge, \cite{yin2020stability} were the first ones to use more sophisticated off-by-one IQCs instead of static multipliers for NN analysis. In this work, in order to further reduce conservatism, we leverage the full potential of dynamic multipliers. More specifically, we combine the general multiplier class of Zames-Falb multipliers \cite{zames1968stability,carrasco2016zames}, the full-block circle criterion and the Yakubovich criterion based on \cite{fetzer2017absolute}. As it is common practice for discrete-time Zames-Falb multipliers \cite{o1967frequency}, we render the IQCs computational via finite impulse response (FIR) filters \cite{wang2014complete,ahmad2014less,carrasco2019convex}.

The main contribution of this paper is the formalization of the use of IQCs for feedback interconnections of LTI systems and NNs, yielding a practical framework for its local stability analysis. We use Lyapunov function arguments to give local stability guarantees and in addition, we provide a method to identify a possibly large inner approximation of the region of attraction (ROA). For our analysis, we establish suitable hard IQC factorizations of a general class of IQCs for slope-restricted nonlinearities, that allow to trade off computational efficiency and conservatism. Finally, we show the improvement over existing methods in numerical examples.

The paper is organized as follows. In Section~\ref{sec:problem_statement}, we present the problem formulation and introduce IQCs for slope-restricted nonlinearities. In Section~\ref{sec:analysis}, we carry out a local  stability analysis using Lyapunov functions and address the computation of ROAs, and in Section~\ref{sec:examples}, we give examples. Finally, in Section~\ref{sec:conclusion} we summarize and conclude the paper.

\textbf{Notation:} By $x\in[a,b]$ for $a,b,x\in\mathbb{R}^n$, we mean that the entries of $x$ are element-wise contained in $[a,b]$ where $a<b$ holds element-wise. We use the notation $\Delta \in [a,b]$, $a,b\in\mathbb{R}^n$ also for diagonal matrices $\Delta = \diag (\delta)$, by which we mean that $\delta\in\mathbb{R}^n$ is contained in $[a,b]$. By an upper index $i$, we mean that the variable $x^i$ belongs to the $i$-th layer of an NN. By $x_j$, we denote the $j$-th component of a vector $x$, except if the index of the vector is $k$, in which case we mean the time index of a sequence of vectors. $\ell_2^n$ denotes the space of square summable functions.

By $\mathbb{S}^{n}$, we denote the set of all symmetric matrices of dimension $n\times n$. By $\mathcal{RH}_\infty$, we mean the space of all causal real-rational transfer functions with poles of absolute value strictly less than 1,
and we denote an ellipsoid around $x_*\in\mathbb{R}^n$ with $X=X^\top\succ 0$ by
\begin{align*}
    \calE(X,x_*) = \{ x \in \bbR^{n} \mid (x - x_*)^\top X (x - x_*) \leq 1 \}.
\end{align*}

\section{NN description using IQCs}\label{sec:problem_statement}
\subsection{Problem formulation}\label{sec:problem_formulation}
We study the discrete-time LTI system
\begin{equation*}\label{eq:LTI}
\text{G}:\hspace{-20mm}
\begin{split}
x_{k+1}&=Ax_k+Bu_k\\
y_k&=Cx_k
\end{split}
\end{equation*}
with state $x_k\in\mathbb{R}^{n_x}$, input $u_k\in\mathbb{R}^{n_u}$, and output $y_k\in\mathbb{R}^{n_y}$. The LTI system is interconnected with an $l$-layer feed-forward NN $u=\mathrm{NN}(y)$ whose input-output behavior $\mathrm{NN}: \mathbb{R}^{n_y}\to \mathbb{R}^{n_u}$ reads
\begin{equation*}\label{eq:NN}
\text{NN}:\hspace{-10mm}
\begin{split}
w^0_k&=y_k\\
v^i_k &= W^{i-1}w^{i-1}_k+b^{i-1},\qquad i=1,\dots,l\\
w^{i}_k&=\phi^{i}(v^i_k),\hspace{24mm} i=1,\dots,l\\
u_k&=W^lw^l_k+b^l.
\end{split}
\hspace{-3mm}
\end{equation*}
Here, $v_k^i\in\mathbb{R}^{n_i}$ and $w_k^i\in\mathbb{R}^{n_i}$ are the inputs and outputs of the neurons, respectively, $W^i\in\mathbb{R}^{n_{i+1}\times n_{i}}$ are the weights, $b^i\in\mathbb{R}^{n_{i+1}}$ are the biases, and $\phi^{i}:\mathbb{R}^{n_i}\to \mathbb{R}^{n_i}$ is the vector of activation functions of the $i$-th layer. 

Our objective is to find an as large as possible inner approximation of the ROA of the interconnection of $\text{G}$ and $\mathrm{NN}$ which we define as follows.

\begin{definition}[Region of attraction]
    Consider a steady state $(x_*,y_*,u_*)$ of the interconnection of $\mathrm{G}$ and $\mathrm{NN}$, i.e., $u_* = \mathrm{NN}(y_*)$, $x_* = A x_* + B u_*$, $y_* = C x_*$. The ROA of this steady state is defined by
    \begin{align*}
        \mathcal{ROA}(x_*) = \{ x_0 \in \bbR^{n_x} \mid \lim_{k\to \infty} x_k = x_* \},
    \end{align*}
    where $x_k$ follows the dynamics $x_{k+1} = A x_k + B \mathrm{NN}(C x_k)$.
\end{definition}
In this paper, we address the following problem, considering ellipsoidal sets $\calE(X,x_*)$ as candidates for inner approximations of $\mathcal{ROA}(x_*)$. 
\begin{problem}
    Find $\calE(X,x_*)$ \emph{as large as possible} such that $\calE(X,x_*) \subseteq \mathcal{ROA}(x_*)$ for the interconnection of $\mathrm{G}$ and~$\mathrm{NN}$.
\end{problem}

As shown in \cite{yin2020stability}, an ROA can be computed using local sector bounds and static multipliers in combination with the specific class of off-by-one IQCs. In this paper, we use similar ideas for a local stability analysis, formalizing the use of general IQCs for the description of the NN. In addition, we propose a systematic approach to find a possibly large $\calE(X,x_*)$.
Throughout the paper, we employ the following notions of local sector and slope restrictions.

\begin{definition}
A function $\varphi: \bbR \to \bbR$ is locally (globally) sector-bounded, $\varphi\in\sec[\alpha,\beta]$, if for all $x\in\mathcal{R}\subset\mathbb{R}$ ($x\in\mathbb{R}$) with $x\neq 0$
\begin{equation}\label{eq:sector_bounds}
\alpha \leq \frac{\varphi(x)}{x} \leq \beta.
\end{equation}
\end{definition}
\begin{definition}
A function $\varphi: \bbR \to \bbR$ is locally (globally) slope-restricted, $\varphi\in\slope[\mu,\nu]$, if for all $x,y\in\mathcal{R}\subset\mathbb{R}$ ($x,y\in\mathbb{R}$) with $x\neq y$
\begin{equation}\label{eq:slope_bounds}
\mu \leq \frac{\varphi(x)-\varphi(y)}{x-y} \leq \nu.
\end{equation}
\end{definition}

Note that the most common activation functions are slope-restricted and sector-bounded, e.g., global slope and sector bounds of ReLU and $\tanh$ are $\alpha_j=\mu_j=0$ and $\beta_j=\nu_j=1$. Yet, the use of local slope and sector bounds reduces conservatism, cf. \cite{yin2020stability}.

We define $\tilde{x}\coloneqq x-x_*$, $\tilde{u}\coloneqq u-u_*$ and $\tilde{y}\coloneqq y-y_*$ for the steady state $(x_*,u_*,y_*)$ and state the incremental dynamics
\begin{equation*}\label{eq:LTI2}
\widetilde{\text{G}}:\hspace{-20mm}
\begin{split}
\tilde{x}_{k+1}&=A\tilde{x}_{k}+B\tilde{u}_k\\
\tilde{y}_k&=C\tilde{x}_k.
\end{split}
\end{equation*}
With the steady state $(v^i_*,w^i_*)$ for all $i=1,\dots,l$ that corresponds to $(x_*,u_*,y_*)$, i.e., $w^0_*=y_*$, $v_*^i=W^{i-1}w_*^{i-1}+b^{i-1}$, $w^i_*=\phi^i(v^i_*)$, $u_*=W^lw^l_*+b^l$,  we define $\tilde{v}^i = v^i - v_*^i$, $\tilde{w}^i = w^i - w_*^i$. By component-wise shifting the activation functions $\phi^i$, we then obtain
\begin{equation*}
\tilde{\phi}^i(\tilde{v}^i)\coloneqq\phi^i(\tilde{v}^i+v_*^i)-\phi^i(v_*^i).
\end{equation*}
Using $\tilde{\phi}^i$, we write the NN in the incremental fashion
\begin{equation*}\label{eq:NN2}
\widetilde{\text{NN}}:\hspace{-14mm}
\begin{split}
\tilde{w}^0_k&=\tilde{y}_k\\
\tilde{v}^i_k &= W^{i-1}\tilde{w}^{i-1}_k,\qquad i=1,\dots,l\\
\tilde{w}^{i}_k&=\tilde{\phi}^{i}(\tilde{v}^i_k),\hspace{13.8mm} i=1,\dots,l\\
\tilde{u}_k&=W^l\tilde{w}^l_k,
\end{split}
\end{equation*}
which conveniently eliminates the bias terms
, while slope-restriction, sector-boundedness and the property $\tilde{\phi}^i(0)=0$ for all $i=1,\dots,l$ are preserved. For activation functions with non-zero offset such as $\mathrm{sigmoid}$ the property $\tilde{\phi}^i(0)=0$ is even obtained through the transformation. Furthermore, we stack up the nonlinearities of all neurons yielding
\begin{equation*}
v=\begin{bmatrix} {v^1}^\top & \dots &  {v^l}^\top \end{bmatrix}^\top,~w=\begin{bmatrix} {w^1}^\top & \dots &  {w^l}^\top \end{bmatrix}^\top\in\mathbb{R}^n
\end{equation*}
with the total number of neurons $n=\sum_{i=1}^l n_{i}$. The variables $\tilde{v}$, $\tilde{w}$, $\tilde{\phi}$, and $\phi$  are defined in the same spirit. Note that the corresponding steady state $(\tilde{x}_*,\tilde{u}_*,\tilde{y}_*,\tilde{v}_*,\tilde{w}_*)$ of $\widetilde{\text{G}}$ and $\widetilde{\text{NN}}$ is at $(0,0,0,0,0)$. The above transformation simplifies the theoretical analysis but it should be noted that, in general, the components $\tilde{\varphi}_j$ of the nonlinearity are not identical and thus, we cannot make use of IQCs for repeated nonlinearities (see Section \ref{sec:tradeoff} for details).

\begin{remark}
Note that for $x_*=0$ and bias-free NNs, i.e., $b^i=0~\forall i=1,\dots, l$, $\mathrm{G}$ corresponds with $\widetilde{\mathrm{G}}$, $\mathrm{NN}$ corresponds with $\widetilde{\mathrm{NN}}$, and $\phi$ corresponds with $\tilde{\phi}$, and the repeated nonlinearities can be exploited in the analysis, yielding less conservative stability certificates.
\end{remark}

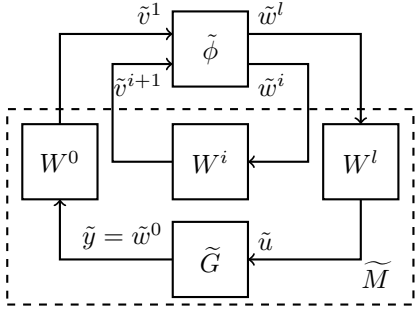
\begin{figure}
\centering
		\begin{tikzpicture}
			\draw[->,thick] (0,-1.5) -- (-0.8,-1.5) -- (-0.8,-0.2) -- (0,-0.2) node[below left ]{$\tilde{v}^{i+1}$} ;
			\draw[->,thick] (1,-0.2) node[below right ]{$\tilde{w}^i$} -- (1.8,-0.2) -- (1.8,-1.5) -- (1,-1.5) ;
			\draw[->,thick] (0,-2.8) node[above left ]{$\tilde{y}=\tilde{w}^0$} -- (-1.5,-2.8) -- (-1.5,-2);
			\draw[->,thick] (-1.5,-1)  -- (-1.5,0.2) -- (0,0.2) node[above left ]{$\tilde{v}^1$};
			\draw[->,thick] (1,0.2) node[above right ]{$\tilde{w}^l$} -- (2.5,0.2) -- (2.5,-1);
			\draw[->,thick] (2.5,-2) -- (2.5,-2.8) -- (1,-2.8)  node[above right ]{$\tilde{u}$};
			\path (0.5,0) node {$\tilde{\phi}$};
			\draw[thick] (0,0.5) rectangle (1,-0.5);
			\path (0.5,-1.5) node {$W^i$};
			\draw[thick] (0,-1) rectangle (1,-2);
			\path (2.5,-1.5) node {$W^l$};
			\draw[thick] (2,-1) rectangle (3,-2);
			\path (-1.5,-1.5) node {$W^0$};
			\draw[thick] (-2,-1) rectangle (-1,-2);
			\path (0.5,-2.8) node {$\widetilde{G}$};
			\draw[thick] (0,-2.3) rectangle (1,-3.3);
			\path (2.7,-3) node {$\widetilde{M}$};
			\draw[thick,dashed] (-2.2,-0.8) rectangle (3.2,-3.4);
		\end{tikzpicture}
		\vspace{-0.2cm}
\caption{Feedback interconnection of plant $\widetilde{G}$ and NN controller.}
\label{fig:Lure}
\end{figure}

To this end, we isolate the diagonal nonlinearities of the activation functions, yielding the feedback interconnection of $\widetilde{M}$ and $\tilde{\phi}$ with signals $\tilde{v}$ and $\tilde{w}$ as shown in Fig.~\ref{fig:Lure}.
For convenience, we use a compact notation similar to \cite{yin2020stability} by defining
\begin{align*}
N_{1:l-1}&\coloneqq\begin{bmatrix} 0 & \cdots & 0 & 0\\ W^1 & \dots & 0 & 0\\ 0 & \ddots &\vdots  & \vdots\\ 0 & \cdots & W^{l-1} & 0 \end{bmatrix}, ~ N_0\coloneqq\begin{bmatrix}W^0C \\ 0\end{bmatrix}.
\end{align*}
to characterize the mapping of the NN by 
\begin{align*}\label{eq:basis_transformations}
\begin{bmatrix}
\tilde{v}\\
\tilde{w}
\end{bmatrix}=
\underbrace{\left[\begin{array}{@{}c|c@{}}
N_0 & N_{1:l-1}  \\
0 & I_{n}
\end{array}\right]}_{\eqqcolon [R_x \vert R_w]}
\begin{bmatrix}
{\tilde{x}}\\
\tilde{w}
\end{bmatrix},~
\tilde{u}=
\underbrace{\begin{bmatrix}
0 & W^l
\end{bmatrix}}_{\eqqcolon R_u}
\tilde{w}.
\end{align*}
The state space realization of $\widetilde{M}$ then is
\begin{equation*}
    \widetilde{M}=
    \left[\begin{array}{@{}c|c@{}}
    A & BR_u\\\hline
    N_0 & N_{1:l-1}
    \end{array}\right].
\end{equation*}

To apply local IQCs, it is necessary to guarantee that the inputs to the neurons $\tilde{v}_k$ stay in some bounded set at all times $k\geq 0$. To identify this set, we bound the inputs to the first layer $\tilde{v}_k^1$ in a box set for a given ROA using energy-to-peak gain techniques from robust control \cite{Palhares2000,grigoriadis1997reduced}. More specifically, similar to \cite{yin2020stability}, we choose an initial symmetric box constraint $[-d^1,d^1]$, $d^1\in\mathbb{R}^{n_1}_+$ for $\tilde{v}_k^1$. From $d^1$, we subsequently determine box constraints for all following layers such that $\tilde{v}_k \in [\udl{d},\ovl{d}] \subseteq \bbR^n$ using interval bound propagation techniques \cite{gowal2018effectiveness}.
Finally, we compute local slope and sector bounds for the activation functions $\tilde{\phi}$ given that $\tilde{v}_k \in [\udl{d},\ovl{d}]$. The following lemma shows how this is accomplished for ReLU activation functions while the same can be achieved for almost any common activation function, see \cite{yin2020stability} for $\tanh$.

\begin{lemma}
    \label{lem:localbounds}
    Assume that $\phi: \bbR^n \to \bbR^n$ is a diagonally repeated ReLU nonlinearity, i.e., $\phi^i(v) = \max\{ 0, v \}$. Then, for the shifted nonlinearity $\tilde{\phi}$ with inputs $\tilde{v}\in[\udl{d},\ovl{d}]$, local slope and sector bounds can be computed as follows: Set for all $i=1,\dots,l$
    \begin{align*}
        \alpha^i &= \frac{\tilde{\phi}^i(\udl{d}^i)}{\udl{d}^i},~\beta^i= \frac{\tilde{\phi}^i (\ovl{d}^i)}{\ovl{d}^i},\\
        \mu^i &= \begin{cases}
        1 & \tilde{\phi}^i(\udl{d}^i) = \udl{d}^i\\
        0 & \text{else}
        \end{cases},~ 
        \nu^i = \begin{cases}
        0 & \tilde{\phi}^i(\ovl{d}^i) = 0\\
        1 & \text{else}
        \end{cases}.
    \end{align*}
    Then $\tilde{\phi}^i \in \slope[\mu^i,\nu^i] \cap \mathrm{sec}[\alpha^i,\beta^i]$.
\end{lemma}

Let us summarize what we have established. We transformed the neural net $\text{NN}$ and system $\text{G}$ with steady state~$(x_*,u_*,y_*)$ to an alternative description $\widetilde{\text{NN}}$ and $\widetilde{\text{G}}$ with steady state~$(0,0,0)$ at the cost of introducing the shifted and no longer repeated nonlinearities $\tilde{\phi}$. Then, we isolated the nonlinearity and finally, we showed how given a box constraint on $\tilde{v}_k^1$, it is possible to compute local slope and sector bounds for $\tilde{\phi}$. The last ingredient which is missing to derive our main result are IQCs for the shifted nonlinearity~$\tilde{\phi}$.

\subsection{Integral quadratic constraints}\label{sec:IQCs}
Within our analysis, we use IQCs \cite{megretski1997system} to describe the nonlinear activation functions that are the elementary components of NNs. In robust control, IQCs are typically used to certify absolute stability under nonlinear model uncertainties. To be consistent with the literature, we first briefly introduce IQCs in the frequency domain and then establish the connection to so-called time domain IQCs, which we use in our stability proof with Lyapunov functions / dissipativity arguments. 
\begin{definition}[Frequency domain IQCs]
Two signals $v\in\ell_2^{n}$ and $w\in\ell_2^{n}$ satisfy the frequency domain IQC defined by $\Pi:\bbC_{|\cdot| = 1}\to\bbC^{2n\times 2n}$ if
\begin{equation*}
    \int_{0}^{2\pi}\begin{bmatrix} \hat{v}(e^{j\omega})\\ \hat{w}(e^{j\omega})\end{bmatrix}^* \Pi(e^{j\omega}) \begin{bmatrix}\hat{v}(e^{j\omega})\\\hat{w}(e^{j\omega})\end{bmatrix} \diff \omega \geq 0,
\end{equation*}
where $\hat{v}(\cdot)$ and $\hat{w}(\cdot)$ are the Fourier transforms of $v$ and $w$.
\end{definition}
In the IQC literature, it is common practice to factorize the multiplier $\Pi(z)$ according to $\Pi(z) = \Psi(z)^* P \Psi(z)$, where $P$ is a symmetric matrix and $\Psi\in\mathcal{RH}_\infty$ is a stable filter. Such a factorization can be used to create a link to the so-called time-domain formulation of IQCs.
\begin{definition}[Time domain IQCs]
With $\Psi\in\mathcal{RH}_\infty$, $P\in\mathbb{R}^{n_r\times n_r}$, let $(\Psi,P)$ be a factorization of $\Pi(z)$. Consider two signals $v \in \ell_2^n$ and $w \in \ell_2^n$ and define the signal $r$ by $r\coloneqq \Psi \begin{bmatrix} v \\ w\end{bmatrix}$. We then say that
 \begin{enumerate}
 \item[(i)] $v,w$ satisfy the time domain \emph{soft} IQC defined by the factorization $(\Psi,P)$ if
 \begin{equation}\label{eq:softIQC}
    \sum_{k=0}^\infty r_k^\top P
    r_k\geq 0.
 \end{equation}
\item[(ii)] $v,w$ satisfy the time domain \emph{hard} IQC defined by $(\Psi,P)$ if, for all $N\in\mathbb{N}_0$,
 \begin{equation}\label{eq:hardIQC}
    \sum_{k=0}^N r_k^\top P
    r_k\geq 0.
 \end{equation}
 \end{enumerate}
Moreover, we say that an operator $\Delta: \ell_2^n \to \ell_2^n$ satisfies the soft/hard IQC defined by $(\Psi,P)$ if \eqref{eq:softIQC}/\eqref{eq:hardIQC} holds for all $v \in \ell_2^n$ and $w = \Delta(v)$.
\end{definition}

Throughout this paper, we focus on the use of hard IQCs.
For the abstraction of NNs, we are primarily interested in IQCs for the sector-bounded, slope-restricted nonlinearity $\tilde{\phi}$. Hence, we provide a collection of IQCs that can be used for describing $\tilde{\phi}$ in the following subsections.

\subsection{Full-block circle / Yakubovich criterion}\label{sec:CY}
While in the current literature mostly static diagonal multipliers are used to describe the sector condition fulfilled by NNs, we suggest the use of dynamic multipliers to significantly reduce conservatism. More specifically, we use the IQCs suggested in \cite{fetzer2017absolute}. From the sector condition \eqref{eq:sector_bounds} one can directly infer
\begin{align}
    \tilde{\phi}(\tilde{v}_k) &= \Delta_k \tilde{v}_k, \label{eq:sector_polytope}
\end{align}
where $\Delta_k = \diag (\delta_k^1,\ldots,\delta_k^n)$ with $\delta_k^j \in [\alpha_j,\beta_j]$, $\delta_k^j = \tilde{\varphi}_j(\tilde{v}_k^j)/\tilde{v}_k^j$.
If we now choose any $\Pi$ from the set
\begin{equation*}
{\bf\Pi}_{[\alpha,\beta]} = \left\{ \Pi\in\mathbb{S}^{2n}\mid \begin{bmatrix}*\end{bmatrix}^\top\Pi\begin{bmatrix}I\\\Delta\end{bmatrix}\succeq 0~\forall\Delta\in[\alpha,\beta]\right\}
\end{equation*}
of \emph{full-block} multipliers for the sector $[\alpha,\beta]$, we can infer
\begin{align}\label{eq:full-block}
    \begin{bmatrix}
    \tilde{v}_k\\
    \tilde{\phi}(\tilde{v}_k)
    \end{bmatrix}^\top
    \Pi
    \begin{bmatrix}
    \tilde{v}_k\\
    \tilde{\phi}(\tilde{v}_k)
    \end{bmatrix} = \tilde{v}_k^\top\begin{bmatrix}
    I\\
    \Delta_k
    \end{bmatrix}^\top
    \Pi
    \begin{bmatrix}
    I\\
    \Delta_k
    \end{bmatrix}\tilde{v}_k\geq 0.
\end{align}
By choosing the factorization $P = \Pi$, $\Psi = I$, we can verify that $\tilde{\phi}$ satisfies the hard IQC defined by $(P,\Psi)$ by simply taking the sum of \eqref{eq:full-block} over $k$. Note that the full-block circle criterion ${\bf\Pi}_{[\alpha,\beta]}$ contains the \emph{diagonal} circle criterion defined by
\begin{equation*}
    {\bf\Pi}_{[\alpha,\beta]}^\mathrm{d}\!=\!\left\{\Pi\in\mathbb{S}^{2n}\!\mid\Pi=\!\begin{bmatrix}
    -2\bar{\alpha}\bar{\beta}\Lambda & (\bar{\alpha}+\bar{\beta})\Lambda\\
    (\bar{\alpha}+\bar{\beta})\Lambda & -2\Lambda
    \end{bmatrix},\Lambda\succeq 0\right\}
\end{equation*}
with $\Lambda=\diag(\lambda)$, $\bar{\alpha}=\diag(\alpha)$ as a special case, $\bar{\beta}=\diag(\beta)$, that coincides with the static multipliers that were used in prior works \cite{fazlyab2020safety,yin2020stability}. 
To also include information about the slope restriction of $\tilde{\phi}$, we notice that \eqref{eq:slope_bounds} implies
\begin{equation*}
    w_{k}-w_{k-1}=\tilde{\phi}(v_{k})-\tilde{\phi}(v_{k-1})=\Delta_k(v_{k}-v_{k-1}).
\end{equation*}
for some diagonal matrices $\Delta_k \in [\mu,\nu]$ for any $k \in \bbN$. This information can now be combined with \eqref{eq:sector_polytope} to infer that any matrix $\Pi$ from the set
\begin{equation*}
{\bf\Pi}^\mathrm{cy} = \left\{ \Pi\in\mathbb{S}^{4n}\mid \begin{bmatrix}*\end{bmatrix}^\top\Pi\begin{bmatrix}I\\\Delta\end{bmatrix}\succeq 0~\forall\Delta\in\left[\begin{bmatrix}
\alpha\\ \mu
\end{bmatrix},\begin{bmatrix}
\beta\\ \nu
\end{bmatrix}\right]\right\}
\end{equation*}
satisfies
\begin{align*}
    \begin{bmatrix}
    \tilde{v}_k\\
    \tilde{v}_k - \tilde{v}_{k-1}\\
    \tilde{\phi}(\tilde{v}_k)\\
    \tilde{\phi}(\tilde{v}_k) - \tilde{\phi}(\tilde{v}_{k-1})
    \end{bmatrix}^\top
    \Pi
    \begin{bmatrix}
    \tilde{v}_k\\
    \tilde{v}_k - \tilde{v}_{k-1}\\
    \tilde{\phi}(\tilde{v}_k)\\
    \tilde{\phi}(\tilde{v}_k) - \tilde{\phi}(\tilde{v}_{k-1})
    \end{bmatrix}&\geq 0 & \forall v \in\ell_{2}^n.
\end{align*}
Consequently, by making use of the time shift operator, we can deduce that $\tilde{\phi}$ satisfies the hard IQC defined by
\begin{align*}
    P\!\in\!{\bf\Pi}^\mathrm{cy} \text{   and   } \Psi^\mathrm{cy}(z)\!=\!
    \begin{bmatrix}
    I & (1-z^{-1})I & 0 & 0\\
    0 & 0 & I & (1-z^{-1})I
    \end{bmatrix}^\top\!.
\end{align*}
Note that this is a dynamic multiplier as the filter $\Psi(z)$ in the factorization depends on the time shift operator $z$ in the time domain or $z \in \bbC_{|\cdot| = 1}$ in the frequency domain. The class $\bf{\Pi}^{\mathrm{cy}}$ is also introduced in \cite{fetzer2017absolute} and related to a time domain version of the so-called \emph{Yakubovich criterion}. Unlike \cite{fetzer2017absolute}, we do not use an additional time shift operator to characterize the circle criterion within the factorization of the combined circle / Yakubovich multipliers above. 

\subsection{Acausal Zames-Falb multipliers}\label{sec:ZF}
In addition to the full-block criteria, we exploit the general class of acausal Zames-Falb multipliers in its discrete time version \cite{Willems1968}. A factorization of causal Zames-Falb IQCs can be found in \cite{lessard2016analysis}.
However, in the following, we provide a factorization of acausal Zames-Falb multipliers, considering FIR multipliers of orders $l_-,l_+$, of the form $H(z)=\sum_{j=l_-}^{l_+}M_j z^{-j}$ with $ M_j\in\mathbb{R}^{n\times n}$ \cite{carrasco2019convex}. 
For ease of exposition, we assume in the following $l_-=l_+=\ell$ and the general case of $l_-\neq l_+$ only requires minor modifications. We start by defining Zames-Falb multipliers like in \cite{fetzer2017absolute} for the sector $[0,\infty]$, which corresponds to monotonicity, by
\begin{equation*}
    {\bf\Pi}_{[0,\infty]}^\mathrm{ZF}\!=\!\left\{\Pi\!\in\!\mathbb{S}^{2n(2\ell\!+1)}\!\mid \Pi\! = \!\begin{bmatrix}
            0 & \widetilde{P}^\top\\
            \widetilde{P} & 0
        \end{bmatrix}\!, \widetilde{P} \! \text{ as in } \! \eqref{eq:pseudoZF} \right\},
\end{equation*}
where the block matrices $\widetilde{P}$ are given by
\begin{align}
    \widetilde{P} = \begin{bmatrix}
            M_0 & M_{-1} & \cdots & M_{-\ell}\\
            M_1 & 0 & \cdots & 0\\
            \vdots & \vdots & \ddots & \vdots\\
            M_\ell & 0 & \cdots & 0
        \end{bmatrix}. \label{eq:pseudoZF}
\end{align}
Herein, the matrices $M_j, j = -\ell,\ldots,\ell$ satisfy
\begin{equation}\label{eq:doubly_hyperdominant}
\sum_{j = -\ell}^\ell M_j e \geq 0,~e^\top \sum_{j = -\ell}^\ell M_j \geq 0
\end{equation}
with the all ones vector $e$, where either all matrix entries except the diagonal of $M_0$ are non-positive or assuming the nonlinearity is odd, the conditions $M_{0,ii}\geq\sum_{k=1,k\neq i}^n \vert M_{0,ik} \vert+ \sum_{j=-\ell}^{\ell}\sum_{k=1}^n \vert M_{j,ik} \vert$, $M_{0,ii}\geq\sum_{k=1,k\neq i}^n \vert M_{0,ki} \vert + \sum_{j=-\ell}^{\ell}\sum_{j=1}^n \vert M_{j,ki} \vert$ hold.
From this class of Zames-Falb multipliers, we obtain the general class ${\bf\Pi}_{[\mu,\nu]}^\mathrm{ZF} = T^\top {\bf\Pi}_{[0,\infty]}^\mathrm{ZF} T$ for the slope bounds $\mu,\nu \in \bbR^n$ by transforming the sector with the matrix
\begin{align*}
    T = \begin{bmatrix}
            I_{2\ell+1} \otimes \diag(\mu) & -I_{ n(2\ell+1)}\\
            -I_{2\ell+1} \otimes \diag(\nu) & I_{ n(2\ell+1) }
        \end{bmatrix}.
\end{align*}
In the following theorem, we present a factorization of the acausal Zames-Falb IQCs that yields a hard IQC, a fact that we exploit in the stability analysis in Section \ref{sec:analysis}.

\begin{theorem}[Hard IQC for acausal Zames Falb]
    \label{thm:ZF}
    Let $\tilde{\phi}: \bbR^n \to \bbR^n$ be a diagonally repeated nonlinear function that is slope-restricted with $\tilde{\phi}\in[\mu,\nu]$ and choose $P \in {\bf\Pi}_{[\mu,\nu]}^\mathrm{ZF}$. Then any sequences $\tilde{v},\tilde{w} \in \ell_2^n$ with $\tilde{w}_k = \tilde{\phi}(\tilde{v}_k)$ satisfy
    \begin{align*}
        \sum_{k=0}^N
        \begin{bmatrix}
            \tilde{v}_k^\top &
            \dots &
            \tilde{v}_{k-\ell}^\top &
            \tilde{w}_k^\top &
            \dots &
            \tilde{w}_{k-\ell}^\top
        \end{bmatrix}
        P
        \begin{bmatrix}
            *
        \end{bmatrix}^\top
        \geq 0,
    \end{align*}
    for all $N \in \bbN_0$, where $\tilde{w}_k,\tilde{v}_k$ are set to zero for $k < 0$, and equivalently, $\tilde{\phi}$ satisfies the hard IQC defined by  $P \in {\bf\Pi}_{[\mu,\nu]}^\mathrm{ZF}$ and
    \begin{align*}
        \Psi^{\mathrm{ZF}}(z) = 
        \begin{bmatrix}
                z^{-0}I & \dots & z^{-\ell}I & 0 & \dots & 0\\
                0 & \dots & 0 & z^{-0}I & \dots & z^{-\ell}I
            \end{bmatrix}^\top.
    \end{align*}
    If $\tilde{\phi}$ is not diagonally repeated, then the statement holds with $M_j = \diag(m_j)$, $m_j \in \bbR^n$.
\end{theorem}
The derivation and proof of this theorem can be found in Appendices \ref{app:derZF} and \ref{app:proofZF}.

\subsection{Trading off conservatism and computational efficiency}\label{sec:tradeoff}
A major difference to previous works is that we use a more general class of dynamic multipliers to describe NNs more accurately. Despite the advantages this provides, it should be noted that such larger classes also increase the computational complexity of the stability test presented in Section \ref{sec:analysis}.
However, a key benefit of the presented approach is its flexibility:
By choosing a smaller class of multipliers, i.e., reducing the number of decision variables, e.g., via the filter order of the Zames-Falb multipliers, we can systematically trade off the computational complexity and the conservatism of our approach.
In particular, the presented class of multipliers contains the static multipliers considered, e.g., in \cite{fazlyab2020safety,pauli2021training,yin2020stability} as special cases, but allows to gradually trade off conservatism and computational efficiency if additional computational resources are available.
In the special case of bias-free NNs, we may significantly reduce conservatism by exploiting the repeated nonlinearities, yielding three versions of Zames-Falb multipliers, respectively for all~$j=-\ell,\dots,\ell$:
\begin{enumerate}
    \item {\bf ZF IQCs}: Diagonal matrices $M_j=\diag(m_j)$.
    \item {\bf ZF-RL IQCs}: Block-diagonal matrices $M_j=\blkdiag(M_j^i), M_j^i\in\mathbb{R}^{n_i\times n_i},i=1,\dots,l$.
    \item {\bf ZF-R IQCs}: Full matrices $M_j$.
\end{enumerate}
Table \ref{tb:multipliers} shows the number of decision variables as well as the required assumptions on the nonlinearity of all previously introduced multipliers.

\begin{table}[H]
\begin{center}
\caption{Decision variables of multipliers}\label{tb:multipliers}
\begin{tabular}{c|ccc}
  & type & nonlinearity & \# dec vars  \\\hline
${\bf\Pi}_{[\alpha,\beta]}^\mathrm{d}$ & sector & general & $n$ \\
${\bf\Pi}_{[\alpha,\beta]}$ & sector & general & $2n(2n+1)/2$ \\
${\bf\Pi}^\mathrm{cy}$ & sector/slope & general & $4n(4n+1)/2$ \\
${\bf\Pi}_{[\mu,\nu]}^\textrm{ZF}$ & slope & general & $(l_-+l_++1)n$ \\
${\bf\Pi}_{[\mu,\nu]}^\textrm{ZF,RL}$ & slope & repeated & $(l_-+l_++1)\sum_{i=1}^ln_i^2$ \\
${\bf\Pi}_{[\mu,\nu]}^\textrm{ZF,R}$ & slope & repeated & $(l_-+l_++1)n^2$ \\\hline
\end{tabular}
\end{center}
\end{table}

A significant benefit of IQCs is that multiple uncertainty descriptions can be combined in a simple way. We can for example combine the full-block multipliers from Section \ref{sec:CY} with the Zames-Falb multipliers from Section \ref{sec:ZF} as it is suggested in \cite{fetzer2017absolute} to obtain the IQC factorization
\begin{align*}
    P \in {\bf\Pi}^\mathrm{comb},~\Psi^\mathrm{comb} = \begin{bmatrix}
    \Psi^\mathrm{ZF}\\
    \Psi^{\mathrm{cy}}
    \end{bmatrix},
\end{align*}
where
\begin{align*}
    {\bf\Pi}^\mathrm{comb}\!=\!\left\{\begin{bmatrix}
            \Pi^\mathrm{ZF} & 0\\
            0 & \Pi^\mathrm{cy}
        \end{bmatrix} \in \bbS \mid \Pi^\mathrm{ZF} \in {\bf\Pi}^\mathrm{ZF}_{[\mu,\nu]}, \Pi^\mathrm{cy} \in {\bf\Pi}^\mathrm{cy} \right\}.
\end{align*}
Alternatively, we can use subclasses of ${\bf\Pi}^\mathrm{comb}$, e.g., combinations of causal or acausal Zames-Falb multipliers with the diagonal or full-block circle criterion that we obtain by replacing $\Pi^\mathrm{cy}$ by $\Pi^\mathrm{c}\in{\bf\Pi}_{[\alpha,\beta]}$ or $\Pi^c\in{\bf\Pi}^\mathrm{d}_{[\alpha,\beta]}$, respectively.
Any diagonal nonlinear function $\tilde{\phi} \in \slope[\mu,\nu] \cap \mathrm{sec}[\alpha,\beta]$, e.g., an activation function of an NN, satisfies the hard IQCs defined by the family of factorizations ${\bf\Pi}^\mathrm{comb}$, $\Psi^\mathrm{comb}$.

\section{Stability analysis}\label{sec:analysis}
In this section, we analyze the proposed feedback interconnection of an LTI system and an NN nonlinearity using the IQC description of NNs introduced in Section \ref{sec:problem_statement}.

\subsection{Local stability analysis}
To give local stability guarantees and an inner approximation of the ROA, we use Lyapunov theory, yielding an LMI certificate for stability similar to \cite{yin2020stability}. 

We choose $\Psi^\mathrm{comb}$ as presented in Section \ref{sec:tradeoff} and find a state space realization
\begin{equation}
\Psi^\mathrm{comb} \begin{bmatrix} \widetilde{M} \\ I\end{bmatrix}=
\left[\begin{array}{@{}c|c@{}}
A_\textrm{tot} & B_\textrm{tot}\\\hline
C_\textrm{tot} & D_\textrm{tot}
\end{array}\right]=
\left[\begin{array}{@{}cc|c@{}}
A & 0 & B R_u\\
B_\psi R_x & A_\psi & B_\psi R_w\\\hline
D_\psi R_x & C_\psi & D_\psi R_w
\end{array}\right],
\label{eq:extendedsys}
\end{equation}
where $A_\psi,B_\psi,C_\psi,D_\psi$ characterize the state space realization of $\Psi^\mathrm{comb}$ and \eqref{eq:extendedsys} in turn describes the dynamical behavior of $\tilde{w}\mapsto r$, where $r$ is the output of $\Psi^\mathrm{comb}$ driven by $\tilde{v}$ and $\tilde{w}$.
With the extended state $\eta^\top=\begin{bmatrix}\tilde{x}^\top & \xi^\top\end{bmatrix}^\top$, where $\xi\in\mathbb{R}^{n_\xi},~n_\xi=2\ell n$, is the state of $\Psi^\mathrm{comb}$, we define the Lyapunov function $V(\eta)=\eta^\top X\eta$ and the ellipsoidal set $\mathcal{E}(X_x,0)=\left\{\tilde{x}\in\mathbb{R}^{n_x}\mid\tilde{x}^\top X_x \tilde{x}\leq 1\right\}$, which is the intersection of the hyperplane $\xi=0$ with the ellipse $\mathcal{E}(X,0)=\left\{\eta\in\mathbb{R}^{n_\eta}\mid\eta^\top X \eta\leq 1\right\}$. While we show stability of the extended state $\eta$, the ROA is characterized by the sector and slope bounds that depend on $\tilde{x}$ only. We define 
$Q\coloneqq W^0C$ and denote the $j$-th row of $Q$ by $Q_j$ and we have $\tilde{v}^1\in[-d^1,d^1]$ with the priorly chosen vector $d^1\in \mathbb{R}^{n_1}_+$. We partition
\begin{equation*}
    X=\begin{bmatrix}
         X_x & X_{x\xi}\\
         X_{\xi x} & X_{\xi}        
    \end{bmatrix}.
\end{equation*}

\begin{theorem} \label{thm:mainThm}
Consider the interconnection of the system $\widetilde{M}$ and the shifted diagonally repeated nonlinearity $\tilde{\phi}$. Assume that $\tilde{\phi} \in \slope[\mu,\nu]\cap \sec[\alpha,\beta]$ for all inputs $\tilde{v} \in [-d,d]$, where $\tilde{v} \in [-d,d]$ is guaranteed by $\tilde{v}^1 \in [-d^1,d^1]$ for the inputs of the first layer. 
Suppose there exist $X = X^\top \succ 0$ and $P\in {\bf\Pi}^\mathrm{comb}$ such that 
\begin{align}\label{eq:LMI}
\left[\begin{array}{@{}cc@{}}
I & 0\\
A_\mathrm{tot} & B_\mathrm{tot}\\\hline
C_\mathrm{tot} & D_\mathrm{tot}
\end{array}\right]^\top\left[\begin{array}{@{}cc|c@{}}
-X & 0 & 0\\
0 & X & 0 \\\hline
0 & 0 & P
\end{array}\right]
\left[\begin{array}{@{}cc@{}}
I & 0\\
A_\mathrm{tot} & B_\mathrm{tot}\\\hline
C_\mathrm{tot} & D_\mathrm{tot}
\end{array}\right]
\prec 0\\\label{eq:invariance}
\begin{bmatrix}
    (d^1_j)^2 & Q_j & 0\\
    Q_j^\top & X_x & X_{x\xi}\\
    0 &X_{\xi x} & X_{\xi}
\end{bmatrix} 
\succeq 0,~j=1,\dots,n_1
\end{align}
hold. Then, for any initial condition $\tilde{x}_0\in\mathcal{E}(X_x,0)$ the feedback interconnection of the LTI system $\widetilde{M}$ and the uncertainty $\tilde{\phi}$ is locally asymptotically stable and $\mathcal{E}(X_x,0)$ is an inner approximation of the ROA.
\end{theorem}
\begin{proof}
    First assume that the slope/sector bounds $\tilde{\phi} \in \slope[\mu,\nu]\cap \sec[\alpha,\beta]$ hold globally (this will later be relaxed).
    Then the sequences $(\tilde{v}_k)$ and $(\tilde{w}_k)$ satisfy the hard IQC defined by $P\in {\bf\Pi}^\mathrm{comb}$ and $\Psi^\mathrm{comb}$.
    As an assumption of the theorem, the LMI \eqref{eq:LMI} is satisfied.
    Multiplying \eqref{eq:LMI} on both sides with $\begin{bmatrix}
            \eta_k^\top & \tilde{w}_k^\top
    \end{bmatrix}^\top$, where $\eta_k$ and $\tilde{w}_k$ are the state and input of the extended system \eqref{eq:extendedsys}, we obtain
    \begin{align}\label{eq:dissi}
        &\begin{bmatrix}
        \eta_k\\ \tilde{w}_k
        \end{bmatrix}^\top\!
        \begin{bmatrix}
        I & 0\\
        A_\mathrm{tot} & B_\mathrm{tot}
        \end{bmatrix}^\top\!
        \begin{bmatrix}
        -X & 0\\
        0 & X
        \end{bmatrix}\!
        \begin{bmatrix}
        I & 0\\
        A_\mathrm{tot} & B_\mathrm{tot}
        \end{bmatrix}\!
        \begin{bmatrix}
        \eta_k\\ \tilde{w}_k
        \end{bmatrix}\!\!\\
        &+
        \begin{bmatrix}
        \eta_k\\ \tilde{w}_k
        \end{bmatrix}^\top\!
        \begin{bmatrix}
        C_\mathrm{tot} & D_\mathrm{tot}
        \end{bmatrix}^\top
        P
        \begin{bmatrix}
        C_\mathrm{tot} & D_\mathrm{tot}
        \end{bmatrix}
        \begin{bmatrix}
        \eta_k\\ \tilde{w}_k
        \end{bmatrix} \leq - \varepsilon \eta_k^\top \eta_k \nonumber 
    \end{align}
    for some small value $\varepsilon > 0$. The term $- \varepsilon \eta_k^\top \eta_k$ can be included due to the strictness of the LMI \eqref{eq:LMI}.
    Since $\eta_k$ is the state of \eqref{eq:extendedsys}, we obtain
    \begin{align*}
        \begin{bmatrix}
        \eta_k\\ \eta_{k+1}
        \end{bmatrix}
        =
        \begin{bmatrix}
        I & 0\\
        A_\mathrm{tot} & B_\mathrm{tot}
        \end{bmatrix}
        \begin{bmatrix}
        \eta_k\\ \tilde{w}_k
        \end{bmatrix},~
        r_k = 
        \begin{bmatrix}
        C_\mathrm{tot} & D_\mathrm{tot}
        \end{bmatrix}
        \begin{bmatrix}
        \eta_k\\ \tilde{w}_k
        \end{bmatrix},
    \end{align*}
    where $r = \Psi^\mathrm{comb} \begin{bmatrix}
            \tilde{v}^\top & \tilde{w}^\top
    \end{bmatrix}^\top$ is the output of the filter $\Psi^\mathrm{comb}$ for two signals with $\tilde{v} = \tilde{\phi}(\tilde{w})$. Plugging this into \eqref{eq:dissi}, yields
    \begin{align*}
        \eta_{k+1}^\top X \eta_{k+1} - \eta_k^\top X \eta_k + r_k^\top P r_k \leq -\varepsilon \eta_k^\top \eta_k.
    \end{align*}
    Summing this inequality from $k = 0,\ldots, N$, we get
    \begin{align*}
        \eta_{N+1}^\top X \eta_{N+1} - \eta_0^\top X \eta_0 + \sum_{k=0}^N r_k^\top P r_k + \sum_{k=0}^N \varepsilon \eta_k^\top \eta_k \leq 0,
    \end{align*}
    where the sum $\sum_{k=0}^N r_k^\top P r_k$ is larger than or equal to zero, since the signals $\tilde{v}$ and $\tilde{w}$ satisfy the IQC defined by $\Psi^\mathrm{comb}$ and $P \in {\bf\Pi}^\mathrm{comb}$. Therefore, we obtain
    \begin{align*}
        \eta_{N+1}^\top X \eta_{N+1} + \sum_{k=0}^N \varepsilon \eta_k^\top \eta_k \leq \eta_0^\top X \eta_0
    \end{align*}
    for all $N \in \bbN_0$, which implies invariance of $\calE(X,0)$ and stability for the extended state $\eta_k$, and that the sum over $\eta_k^\top \eta_k$ is bounded for any $N\in\mathbb{N}$. In combination, these facts also imply asymptotic stability of $\eta = 0$ and thus of $\tilde{x} = 0$. 
    %
    As a last step, we estimate the peak of the activations in $\tilde{v}_k^1$ as follows
    \begin{align*}
        (Q_j \tilde{x}_k)^2 = \tilde{x}_k^\top Q_j^\top Q_j \tilde{x}_k = \eta_k^\top \begin{bmatrix}
        Q_j^\top \\
        0
        \end{bmatrix}^\top 
        \begin{bmatrix}
        Q_j^\top \\
        0
        \end{bmatrix}
        \eta_k.
    \end{align*}
    Combining this with
    \begin{align*}
        \begin{bmatrix}
        Q_j^\top \\
        0
        \end{bmatrix}
        (d^1_j)^{-2}
        \begin{bmatrix}
        Q_j^\top \\
        0
        \end{bmatrix}^\top
        \preceq X,
    \end{align*}
    which is obtained by a Schur complement of \eqref{eq:invariance}, yields
    \begin{align*}
        (Q_j \tilde{x}_k)^2 \leq (d^1_j)^2 \eta_k^\top X\eta_k \leq (d^1_j)^2 \eta_0^\top X \eta_0.
    \end{align*}
    Hence, if $\eta_0$ is initialized in $\calE(X,0)$ (or $\tilde{x}_0$ is initialized in $\calE(X_x,0)$ and $\xi_0$ is set to zero), then $\eta_0^\top X \eta_0 \leq 1$ and therefore $Q_j \tilde{x}_k$ is bounded by $d_j^1$, i.e., $\vert Q_j\tilde{x}_k\vert\leq d_j^1$ at all times $k\geq 0$. 
    Hence, the behavior of $\tilde{\phi}$ outside of $[-d,d]$ is irrelevant if $\tilde{x}_0 \in \calE(X_x,0)$ and therefore, the results of this theorem persist to hold when $\tilde{\phi}$ satisfies the sector and slope bounds only locally in $[-d,d]$.
\end{proof}

\begin{corollary}\label{cor:GandNN}
    If the conditions of Theorem \ref{thm:mainThm} are satisfied, then $\calE(X_x,x_*)$ is an inner approximation of the ROA for $x_*$ of the original interconnection of $\text{G}$ and $\text{NN}$.
\end{corollary}
Corollary \ref{cor:GandNN} allows to verify stability and compute an ROA for the linear system $\text{G}$ with NN component $\text{NN}$ by computing a positive semidefinite $X$ that renders \eqref{eq:LMI} and \eqref{eq:invariance} feasible. Note that Theorem 1 in \cite{yin2020stability} is a special case of Corollary \ref{cor:GandNN} that we adapted to the use of general IQCs.
\subsection{Computing regions of attraction}\label{sec:ROA}
To obtain a possibly large inner approximation $\calE (X_x,x_*)$ of the ROA, we suggest the minimization of the trace of $X_x$ to obtain a matrix $X_x$ with small eigenvalues (which corresponds to a large ellipsoid), subject to \eqref{eq:LMI} and \eqref{eq:invariance}. We must simultaneously optimize over $d^1 \in \bbR^{n_1}$, $X_x$ and $P \in {\bf\Pi}^\mathrm{comb}$. For a given value of $d^1$, we can solve the semidefinite program (SDP)
    \begin{equation}\label{eq:SDP}
        \min_{X\succ 0,P \in {\bf\Pi}^\mathrm{comb}} \trace{X_x}\quad\text{s.\,t.}~\eqref{eq:LMI}~\text{and}~ \eqref{eq:invariance},
    \end{equation}
to identify the largest inner approximation of the ROA. 
We note that \eqref{eq:LMI} is not linear in $d^1$ as the slope and sector bounds depend on $\udl{d},\ovl{d}$ that are determined from $d^1$, cf. Section \ref{sec:problem_formulation}. To additionally optimize over $d^1$, we use bisection and convex linesearch methods as follows.

First, we use a bisection method to determine the maximum value of $d^1$ for which the SDP \eqref{eq:SDP} is feasible and consequently, we run a golden section search to find the minimum value of $\trace(X_x)$ over $\delta$ with $d^1=\delta \cdot e_n, \delta>0$, where $e_n$ is the all ones vector of dimension $n$. There is an underlying trade-off that establishes this minimum. While small values of $d^1$ constrain the inner approximation of the ROA to be within those given bounds, for large values of $d^1$ the more conservative slope and sector bounds are the limiting factor.

\section{Numerical examples}\label{sec:examples}
In this section, we illustrate the applicability of our approach. We use YALMIP \cite{Lofberg2004} in combination with Mosek \cite{mosek} to solve the SDP \eqref{eq:SDP}\footnote{The code is available at https://github.com/ppauli/IQCs-for-NNs}.

\subsection{Exploiting repeated nonlinearities}
In the following, we analyze the dynamics of an inverted pendulum linearized in the upright position, i.e.,
\begin{equation}\label{eq:inv_pendulum}
\dot{x}=
    \begin{bmatrix}0 & 1\\
    \frac{g}{L}& -\frac{\mu}{mL^2}\end{bmatrix}x+
    \begin{bmatrix}0 \\
    \frac{1}{mL^2}\end{bmatrix}u
\end{equation}
with mass $m=0.15\,\text{kg}$, length $L=0.5\,\text{m}$, friction coefficient $\mu=0.5\,\text{Nms/rad}$ and gravitational acceleration $g=9.81\,\text{N}/\text{kg}$. The plant state $x=[\theta,\dot{\theta}]$ consists of the angle and the angular velocity of the pendulum. We discretize the dynamics with $dt=0.02\,\textrm{s}$ and the resulting discrete-time LTI system is interconnected with a bias-free NN controller $u=\text{NN}(x)$ with $n_1=n_2=5$ and activation function $\tanh$. We train this NN from input-state data obtained from an MPC with input constraint $\vert u \vert\leq1$, that stabilizes the linearized inverted pendulum with dynamics \eqref{eq:inv_pendulum}.

\begin{figure}
    \centering
    \includegraphics[scale=0.9]{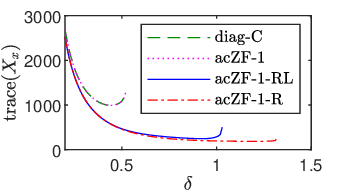} \includegraphics[scale=0.9]{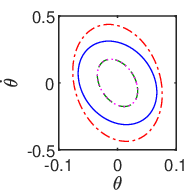} 
    \caption{Comparison of diagonal circle criterion {\small ${\bf\Pi}_{[\alpha,\beta]}^\mathrm{d}$} combined with diagonal ${\bf\Pi}_{[\mu,\nu]}^\textrm{ZF}$, block-diagonal ${\bf\Pi}_{[\mu,\nu]}^\textrm{ZF,RL}$ and full-block ${\bf\Pi}_{[\mu,\nu]}^\textrm{ZF,R}$ acausal Zames-Falb multipliers  with $\ell=1$ for linearized inverted pendulum.}
    \label{fig:traceX}
\end{figure}

In the following, we compare the solution of \eqref{eq:SDP}, which characterizes the size of the inner approximation of the ROA, for different values of $\delta$. We describe the NN using different classes of multipliers with varying complexity. In particular, we employ the diagonal circle criterion (diag-C) 
as well as the diagonal circle criterion in combination with acausal Zames-Falb multipliers of filter order $\ell=1$ for all three kinds of multipliers presented in Section \ref{sec:tradeoff}. Fig.~\ref{fig:traceX} shows the resulting values of $\trace(X_x)$ over $\delta$ and the ellipses at the corresponding minima of $\trace(X_x)$ for the different choices of $M_j$. We notice that in this example, the use of Zames-Falb multipliers with diagonal $M_j$ (acZF-1) has hardly noticeable benefits to the circle criterion (diag-C). However, block-diagonal (acZF-1-RL) and full-block (acZF-1-R) multipliers $M_j$, exploiting the repeated nonlinearity in the slope restriction condition of the bias-free NN, yield significantly smaller values for $\trace(X_x)$ which corresponds to larger guaranteed ROAs. 

\subsection{Benefits of Zames-Falb multipliers}
In the following, we adopt an example from \cite{yin2020stability} on lateral vehicle control (see \cite{yin2020stability} for the parameters). The LTI system has the form
\begin{equation*}
    x_{k+1}=Ax_k+B_1u_{\textrm{sat},k}+B_2q_k.
\end{equation*}
The feedback interconnection contains a saturation block $u_\textrm{sat}=\textrm{sat}(u)$ acting on the output $u$ of the NN, with saturation limit $u_\textrm{max}=\pi/6$, and a norm-bounded LTI uncertainty $\Delta_\textrm{LTI}\in\mathcal{RH}_\infty$ with $\Vert\Delta_\mathrm{LTI}\Vert\leq 0.1$, such that $q=\Delta_\mathrm{LTI}(u_\textrm{sat})$. Both additional nonlinearities can be captured by quadratic constraints which can be included in the condition \eqref{eq:LMI} using standard methods, cf. \cite{veenman2016robust}. We use the state feedback NN controller obtained by \cite{yin2020stability} using policy gradient methods, that is an 2-layer NN with $n_1=n_2=32$ and activation function $\tanh$.

To describe the NN, we use the circle criterion (diag-C), causal Zames-Falb multipliers (diag-cZF) and acausal Zames multipliers (diag-acZF), both with filter order $\ell=1$. For all multipliers, we proceed according to Section \ref{sec:ROA} to identify a large inner approximation of the ROA. We perform bisection over $\delta$ to retrieve the maximum $\delta_{\max}$ such that \eqref{eq:SDP} is feasible and subsequently, golden sectioning to find the minimum of \eqref{eq:SDP} over $\delta$. Table \ref{tb:ZF} compares the resulting minimum $\trace(X_x)$, the maximum feasible value $\delta_{\max}$, the number of decision variables, and the required computation time to solve \eqref{eq:SDP} on a standard i7 note book. Using dynamic Zames-Falb multipliers (diag-cZF, diag-acZF) results in much lower values of $\trace(X_x)$ in comparison to static multipliers (diag-C). Note that the causal Zames-Falb multipliers (diag-cZF) coincide with the off-by-one IQCs used in \cite{yin2020stability}. While \cite{yin2020stability} report $\trace (X_x)=2.9$ at $\delta=0.6$, the systematic procedure presented in Section \ref{sec:ROA} returns even smaller values of $\trace (X_x)$. Using acausal Zames-Falb multipliers (diag-acZF) instead of causal ones (diag-aZF), this value can be improved further, i.e., yields an even larger inner approximation of the ROA.
\vspace{-0.0cm}
\begin{table}[H]
\begin{center}
\caption{Benefits of Zames-Falb multipliers}\label{tb:ZF}
\begin{tabular}{c|cccc}
  & $\min_{\delta} \trace(X_x)$& $\delta_{\max}$ & \# dec vars & Com time \\\hline
diag-C & $3.842$ & $0.67$ & $98$ & $0.8785\,\mathrm{s}$\\
diag-cZF & $2.726$ & $1.47$ & $2,754$ & $123.9\,\mathrm{s}$\\
diag-acZF & $2.696$ & $1.51$ & $9,442$ & $2095\,\mathrm{s}$ \\\hline
\end{tabular}
\vspace{-0.3cm}
\end{center}
\end{table}

\section{Conclusion}\label{sec:conclusion}
In this paper, we studied feedback interconnections of an LTI system with an NN nonlinearity in discrete time and analayzed local stability thereof. We used IQCs to describe NNs, exploiting the sector-bounded and slope-restricted structure of the underlying activation functions. In contrast to existing approaches, we leveraged the full potential of dynamic IQCs to describe the nonlinear activation functions in a less conservative fashion. We considered a number of IQCs for slope-restricted nonlinearites, including acausal Zames-Falb multipliers, and derived LMI based stability certificates. Through the choice of IQCs and multipliers we traded of computational complexity and conservatism. In addition, we discussed how an inner-approximation of the corresponding ROA can be computed.

In future research, we plan to explore more scalable relaxations of full-block multipliers and their applicability to the proposed analysis.





\bibliographystyle{IEEEtran}
\bibliography{references}

\begin{thebibliography}{10}
\providecommand{\url}[1]{#1}
\csname url@samestyle\endcsname
\providecommand{\newblock}{\relax}
\providecommand{\bibinfo}[2]{#2}
\providecommand{\BIBentrySTDinterwordspacing}{\spaceskip=0pt\relax}
\providecommand{\BIBentryALTinterwordstretchfactor}{4}
\providecommand{\BIBentryALTinterwordspacing}{\spaceskip=\fontdimen2\font plus
\BIBentryALTinterwordstretchfactor\fontdimen3\font minus
  \fontdimen4\font\relax}
\providecommand{\BIBforeignlanguage}[2]{{%
\expandafter\ifx\csname l@#1\endcsname\relax
\typeout{** WARNING: IEEEtran.bst: No hyphenation pattern has been}%
\typeout{** loaded for the language `#1'. Using the pattern for}%
\typeout{** the default language instead.}%
\else
\language=\csname l@#1\endcsname
\fi
#2}}
\providecommand{\BIBdecl}{\relax}
\BIBdecl

\bibitem{fazlyab2020safety}
M.~Fazlyab, M.~Morari, and G.~J. Pappas, ``Safety verification and robustness
  analysis of neural networks via quadratic constraints and semidefinite
  programming,'' \emph{IEEE Trans. Automat. Contr.}, 2020.

\bibitem{hashemi2021certifying}
N.~Hashemi, J.~Ruths, and M.~Fazlyab, ``Certifying incremental quadratic
  constraints for neural networks via convex optimization,'' in \emph{Learning
  for Dynamics and Control}.\hskip 1em plus 0.5em minus 0.4em\relax PMLR, 2021,
  pp. 842--853.

\bibitem{levin1993control}
A.~U. Levin and K.~S. Narendra, ``Control of nonlinear dynamical systems using
  neural networks: Controllability and stabilization,'' \emph{IEEE Transactions
  on neural networks}, vol.~4, no.~2, pp. 192--206, 1993.

\bibitem{suykens1995artificial}
J.~A. Suykens, J.~P. Vandewalle, and B.~L. de~Moor, \emph{Artificial neural
  networks for modelling and control of non-linear systems}.\hskip 1em plus
  0.5em minus 0.4em\relax Springer Science \& Business Media, 1995.

\bibitem{fazlyab2019efficient}
M.~Fazlyab, A.~Robey, H.~Hassani, M.~Morari, and G.~Pappas, ``Efficient and
  accurate estimation of {L}ipschitz constants for deep neural networks,'' in
  \emph{Advances in Neural Information Processing Systems}, 2019, pp.
  11\,423--11\,434.

\bibitem{pauli2021training}
P.~Pauli, A.~Koch, J.~Berberich, P.~Kohler, and F.~Allg{\"o}wer, ``Training
  robust neural networks using {L}ipschitz bounds,'' \emph{IEEE Control Syst.
  Lett.}, 2021.

\bibitem{yin2020stability}
H.~Yin, P.~Seiler, and M.~Arcak, ``Stability analysis using quadratic
  constraints for systems with neural network controllers,'' \emph{arXiv
  preprint arXiv:2006.07579}, 2020.

\bibitem{pauli2020offset}
P.~Pauli, J.~K{\"o}hler, J.~Berberich, A.~Koch, and F.~Allg{\"o}wer,
  ``Offset-free setpoint tracking using neural network controllers,''
  \emph{arXiv preprint arXiv:2011.14006}, 2020.

\bibitem{revay2020convex}
M.~Revay, R.~Wang, and I.~R. Manchester, ``A convex parameterization of robust
  recurrent neural networks,'' \emph{IEEE Control Syst. Lett.}, vol.~5, no.~4,
  pp. 1363--1368, 2020.

\bibitem{megretski1997system}
A.~Megretski and A.~Rantzer, ``System analysis via integral quadratic
  constraints,'' \emph{IEEE Trans. Automat. Contr.}, vol.~42, no.~6, pp.
  819--830, 1997.

\bibitem{veenman2016robust}
J.~Veenman, C.~W. Scherer, and H.~K{\"o}ro{\u{g}}lu, ``Robust stability and
  performance analysis based on integral quadratic constraints,''
  \emph{European Journal of Control}, vol.~31, pp. 1--32, 2016.

\bibitem{gonzaga2012stability}
C.~A. Gonzaga, M.~Jungers, and J.~Daafouz, ``Stability analysis of
  discrete-time {Lur’e} systems,'' \emph{Automatica}, vol.~48, no.~9, pp.
  2277--2283, 2012.

\bibitem{ahmad2012lmi}
N.~S. Ahmad, W.~P. Heath, and G.~Li, ``{LMI}-based stability criteria for
  discrete-time {L}ur'e systems with monotonic, sector-and slope-restricted
  nonlinearities,'' \emph{IEEE Trans. Automat. Contr.}, vol.~58, no.~2, pp.
  459--465, 2012.

\bibitem{zames1968stability}
G.~Zames and P.~Falb, ``Stability conditions for systems with monotone and
  slope-restricted nonlinearities,'' \emph{SIAM Journal on Control}, vol.~6,
  no.~1, pp. 89--108, 1968.

\bibitem{carrasco2016zames}
J.~Carrasco, M.~C. Turner, and W.~P. Heath, ``{Z}ames--{F}alb multipliers for
  absolute stability: From {O' Shea' s} contribution to convex searches,''
  \emph{European Journal of Control}, vol.~28, pp. 1--19, 2016.

\bibitem{fetzer2017absolute}
M.~Fetzer and C.~W. Scherer, ``Absolute stability analysis of discrete time
  feedback interconnections,'' \emph{IFAC-PapersOnLine}, vol.~50, no.~1, pp.
  8447--8453, 2017.

\bibitem{o1967frequency}
R.~O'Shea and M.~Younis, ``A frequency-time domain stability criterion for
  sampled-data systems,'' \emph{IEEE Trans. Automat. Contr.}, vol.~12, no.~6,
  pp. 719--724, 1967.

\bibitem{wang2014complete}
S.~Wang, W.~P. Heath, and J.~Carrasco, ``A complete and convex search for
  discrete-time noncausal {FIR} {Z}ames-{F}alb multipliers,'' in \emph{53rd
  IEEE Conference on Decision and Control}.\hskip 1em plus 0.5em minus
  0.4em\relax IEEE, 2014, pp. 3918--3923.

\bibitem{ahmad2014less}
N.~S. Ahmad, J.~Carrasco, and W.~P. Heath, ``A less conservative {LMI}
  condition for stability of discrete-time systems with slope-restricted
  nonlinearities,'' \emph{IEEE Trans. Automat. Contr.}, vol.~60, no.~6, pp.
  1692--1697, 2014.

\bibitem{carrasco2019convex}
J.~Carrasco, W.~P. Heath, J.~Zhang, N.~S. Ahmad, and S.~Wang, ``Convex searches
  for discrete-time {Z}ames--{F}alb multipliers,'' \emph{IEEE Trans. Automat.
  Contr.}, vol.~65, no.~11, pp. 4538--4553, 2019.

\bibitem{Palhares2000}
R.~M. Palhares and P.~L. Peres, ``Robust filtering with guaranteed
  energy-to-peak performance an {LMI} approach,'' \emph{Automatica}, vol.~36,
  no.~6, pp. 851--858, 2000.

\bibitem{grigoriadis1997reduced}
K.~M. Grigoriadis and J.~T. Watson, ``Reduced-order {$H_\infty$} and
  {$L_2$--$L_\infty$} filtering via linear matrix inequalities,'' \emph{IEEE
  Trans. Aerosp. Electron. Syst.}, vol.~33, no.~4, pp. 1326--1338, 1997.

\bibitem{gowal2018effectiveness}
S.~Gowal, K.~Dvijotham, R.~Stanforth, R.~Bunel, C.~Qin, J.~Uesato,
  R.~Arandjelovic, T.~Mann, and P.~Kohli, ``On the effectiveness of interval
  bound propagation for training verifiably robust models,'' \emph{arXiv
  preprint arXiv:1810.12715}, 2018.

\bibitem{Willems1968}
J.~Willems and R.~Brockett, ``Some new rearrangement inequalities having
  application in stability analysis,'' \emph{{IEEE} Transactions on Automatic
  Control}, vol.~13, no.~5, pp. 539--549, 1968.

\bibitem{lessard2016analysis}
L.~Lessard, B.~Recht, and A.~Packard, ``Analysis and design of optimization
  algorithms via integral quadratic constraints,'' \emph{SIAM Journal on
  Optimization}, vol.~26, no.~1, pp. 57--95, 2016.

\bibitem{Lofberg2004}
J.~L{\"{o}}fberg, ``Yalmip : A toolbox for modeling and optimization in
  matlab,'' in \emph{In Proc. of the CACSD Conference}, Taipei, Taiwan, 2004.

\bibitem{mosek}
\BIBentryALTinterwordspacing
{MOSEK ApS}, \emph{The MOSEK optimization toolbox for MATLAB manual. Version
  9.0.}, 2019. [Online]. Available:
  \url{http://docs.mosek.com/9.0/toolbox/index.html}
\BIBentrySTDinterwordspacing

\bibitem{DAmato2001}
F.~D'Amato, M.~Rotea, A.~Megretski, and U.~Jönsson, ``New results for analysis
  of systems with repeated nonlinearities,'' \emph{Automatica}, vol.~37, no.~5,
  pp. 739--747, 2001.

\end{thebibliography}






\appendix

\subsection{Derivation of Theorem \ref{thm:ZF}} \label{app:derZF}

The derivation of Theorem \ref{thm:ZF} is inspired by \cite{fetzer2017absolute} making use of doubly hyperdominant matrices.

\begin{definition}[Doubly hyperdominant matrix]
    A matrix $M \in \bbR^{n\times n}$ is called doubly hyperdominant if the matrix is diagonally dominant with non-positive off-diagonal terms and non-negative diagonal ones, i.e.,
    \begin{align*}
        (M)_{ij} \leq 0 ~\forall i\neq j \text{ and } e^\top M \geq 0, M e \geq 0.
    \end{align*}
\end{definition}
We also use the following lemma from \cite{fetzer2017absolute,Willems1968}.
\begin{lemma}
    \label{lem:PreZF}
    Let $\varphi$ be slope-restricted in $[0,\infty]$ and $M \in \bbR^{n\times n}$ be doubly hyperdominant. Then the diagonally repeated function $\phi: \bbR^n \to \bbR^n, (v_1,\ldots,v_n) \mapsto (\varphi (v_1),\ldots,\varphi(v_n))$ satisfies for all $v \in \bbR^n$
    \begin{align*}
        v^\top M \phi(v) \geq 0.
    \end{align*}
\end{lemma}

\subsection{Proof of Theorem \ref{thm:ZF}}\label{app:proofZF}

First, consider the case $[\mu,\nu] = [0,\infty]$ and assume that $\tilde{\phi}$ is diagonally repeated. The following matrix
\begin{align*}
    M = \begin{bmatrix}
        M_0 & \cdots & M_{-\ell} & 0 & \cdots & 0\\
        \vdots & M_0 & \cdots & M_{-\ell} & \ddots & \vdots \\
        M_{\ell} & \vdots & \ddots & & \ddots & 0\\
        0 & M_{\ell} & & & & M_{\ell}\\
        \vdots & \ddots & \ddots  & & \ddots & \vdots \\
        0 & \cdots & 0 & M_{\ell} & \cdots & M_0
    \end{bmatrix} \label{eq:M}
\end{align*}
is doubly hyperdominant by \eqref{eq:doubly_hyperdominant} and assumptions on $M_{-\ell},\ldots,M_{\ell}$. Hence, we obtain from Lemma~\ref{lem:PreZF} that
\begin{align*}
    \begin{bmatrix}
        v_0\\ \vdots \\ v_N
    \end{bmatrix}^\top\!
    \hspace{-2mm}
    M
    \begin{bmatrix}
        w_0\\ \vdots \\ w_N
    \end{bmatrix}
    &\!=\!
    \sum_{k = 0}^N
    \underbrace{\begin{bmatrix}
        v_k\\ \vdots \\ v_{k-\ell}
    \end{bmatrix}^\top
    \hspace{-1mm}}_{\eqqcolon\udl{v}_k^\top}\!
    \underbrace{
        \begin{bmatrix}
         M_0 & \cdots & M_{-\ell} \\
         \vdots & \multicolumn{2}{c}{\vspace{0.2cm}\multirow{2}{*}{\hspace{-0.3cm} \tiny 
         $\begin{array}{ccc}
    0 & \dots & 0\\
    \vdots & \ddots & \vdots\\
    0 & \dots & 0\\   
    \end{array}$}}  \\
         M_{\ell} &  
    \end{bmatrix}
    }_{ = \widetilde{P}}\!
    \underbrace{\begin{bmatrix}
        w_k\\ \vdots \\ w_{k-\ell}
    \end{bmatrix}}_{\eqqcolon\udl{w}_k}\\
    &=
    \frac{1}{2}
    \sum_{k = 0}^N
    \begin{bmatrix}
    \udl{v}_k\\
    \udl{w}_k
    \end{bmatrix}^\top
    P
    \begin{bmatrix}
    \udl{v}_k\\
    \udl{w}_k
    \end{bmatrix}\geq 0,
\end{align*}
where $v_1,\ldots,v_N \in \bbR^n, w_1,\ldots, w_N \in \bbR^n$ are sequences with $w_k = \phi (v_k) $ for $k = 1, \ldots ,N$ and $v_{-1},\ldots,v_{-\ell}$ and $w_{-1},\ldots,w_{-\ell}$ are set to zero. This proves the time domain version of Theorem \ref{thm:ZF} for the case $[\mu,\nu] = [0,\infty]$ and repeated nonlinearities. The frequency domain version follows directly. The generalization to non-repeated nonlinearities and slopes restricted to different sectors are discussed below.
\begin{itemize}
    \item \emph{Non-repeated nonlinearities:} The result for non-repeated nonlinearities can easily be obtained by constraining $M_j = \diag(m_j), j = -\ell,\ldots,\ell$ as in \cite{fetzer2017absolute}.
    \item \emph{Generalization to slope-restricted nonlinearities:} In Theorem \ref{thm:ZF} the matrix $T$ is used to generalize the result to other sectors. This matrix transforms a function $\varphi$ with slope in $[\mu,\nu]$ to a multi-valued mapping $\tilde{\varphi}$ with slope in $[0,\infty]$. This is the principle that allows the generalization to other sectors, cf. \cite{DAmato2001}. \qed
\end{itemize}

\end{document}